# Direct observation of 3D topological spin textures and their interactions using soft x-ray vector ptychography

Arjun Rana[1,2*], Chen-Ting Liao[2,3*], Ezio Iacocca[4], Ji Zou[1], Minh Pham[2,5], Emma-Elizabeth Cating Subramanian[2,3], Yuan Hung Lo[1,2], Sinéad A. Ryan[2,3], Xingyuan Lu[1,2], Charles S. Bevis[2,3], Robert M. Karl Jr[2,3], Andrew J. Glaid[6], Young-Sang Yu[7], Pratibha Mahale[6], David A. Shapiro[7], Sadegh Yazdi[8], Thomas E. Mallouk[6], Stanley J. Osher[5], Henry C. Kapteyn[2,3], Vincent H. Crespi[6], John V. Badding[6], Yaroslav Tserkovnyak[1], Margaret M. Murnane[2,3], Jianwei Miao[1,2†]

[1]Department of Physics & Astronomy and California NanoSystems Institute, University of California, Los Angeles, CA 90095, USA. [2]STROBE Science and Technology Center. [3]JILA and Department of Physics, University of Colorado and NIST, 440 UCB, Boulder, Colorado 80309, USA. [4]Department of Mathematics, Physics, and Electrical Engineering, Northumbria University, Newcastle upon Tyne, NE1 8ST, UK. [5]Department of Mathematics, University of California, Los Angeles, CA 90095, USA. [6]Departments of Chemistry, Physics, Materials Science and Engineering and Materials Research Institute, Penn State University, University Park, PA 16802, USA. [7]Advanced Light Source, Lawrence Berkeley National Laboratory, Berkeley, CA 94720, USA. [8]Renewable and Sustainable Energy Institute, University of Colorado, Boulder, CO 80309, USA.

[*]These authors contributed equally to this work. [†]Correspondence author. Email: miao@physics.ucla.edu (J.M.)

**Magnetic topological defects are energetically stable spin configurations characterized by symmetry breaking. Vortices and skyrmions are two well-known examples of 2D spin textures that have been actively studied for both fundamental interest and practical applications. However, experimental evidence of the 3D spin textures has been largely indirect or qualitative to date, due to the difficulty of quantitatively characterizing them within**

**nanoscale volumes. Here, we develop soft x-ray vector ptychography to quantitatively image the 3D magnetization vector field in a frustrated superlattice with 10 nm spatial resolution. By applying homotopy theory to the experimental data, we quantify the topological charge of hedgehogs and anti-hedgehogs as emergent magnetic monopoles and probe their interactions inside the frustrated superlattice. We also directly observe virtual hedgehogs and anti-hedgehogs created by magnetically inert voids. We expect that this new quantitative imaging method will open the door to study 3D topological spin textures in a broad class of magnetic materials. Our work also demonstrates that magnetically frustrated superlattices could be used as a new platform to investigate hedgehog interactions and dynamics and to exploit optimized geometries for information storage and transport applications.**

Coherent diffractive imaging, which replaces the lens of a microscope with a computational algorithm (*1*), is revolutionizing the imaging and microscopy field (*2*). In particular, ptychography, a powerful scanning coherent diffractive imaging method, has found broad applications using synchrotron radiation, high harmonic generation, electron and optical microscopy (*3-10*). By employing circularly polarized hard x-rays, ptychography has been combined with tomography to image the three-dimensional (3D) magnetization vector field of a $GdCo_2$ pillar sample with a spatial resolution of 100 nm (*11*). However, soft x-rays have two unique advantages over hard x-rays for magnetic coherent diffractive imaging: an optimal magnetic contrast at the L-edge resonance of transition metals, and an increase of the scattering cross section by about two orders of magnitude (*12-14*). Here, we developed 3D soft x-ray vector ptychography for the simultaneous reconstruction of the electron density and magnetization vector field with a spatial resolution of 10 nm. The high-resolution and quantitative nature of our new method enabled us to invoke homotopy theory (*15*) and quantify the topological charge of 3D spin textures directly from experimental data. This represents a significant advantage over previous methods that either rely on *a priori* assumptions (*16*) or use Maxwell's equations as a constraint (*17, 18*) to obtain the 3D vector field.

We utilized soft x-ray vector ptychography to quantitatively image the 3D magnetization vector field of a ferromagnetic superlattice (*19, 20*). The geometric constraint in the superlattice competes with the exchange stiffness and magnetostatic energy to produce a magnetically frustrated configuration that harbors non-trivial spin textures. We identified a large number of 3D magnetic defects in the superlattice and directly quantified their topological charge as hedgehogs and anti-hedgehogs at room temperature, in the absence of an external magnetic field. As emergent magnetic monopoles (*21*), hedgehogs can create, annihilate and manipulate skyrmions (*22, 23*). Compared to local 2D topological objects such as skyrmions (*24, 25*), hedgehogs are not only topologically protected, but also are robust against thermal fluctuations due to their nonlocality. Thus, hedgehogs could be used as more stable topological carriers than skyrmions for magnetic information storage and processing (*26*).

The experiment was performed using circularly polarized soft x-rays that were focused onto a superlattice sample using a Fresnel zone plate (Fig. 1). The sample was fabricated by infiltrating nickel into a face-centered cubic superlattice of silica nanospheres with a diameter of 60 nm to form an intricate ferromagnetic network (*20*) (see supplementary materials). The resulting ferromagnetic superlattice consists of octahedral and tetrahedral sites connected by thin filaments. The magnetic contrast of the sample was obtained by using x-ray magnetic circular dichroism (*12-14*) and tuning the x-ray energy to the L-edge of nickel (856 eV). To separate the charge density from the magnetization vector field, two independent measurements were made by using left- and right-circularly polarized soft x-rays. In each measurement, three independent tilt series were collected from the sample, corresponding to three in-plane rotation angles (0°, 120° and 240°) around the z-axis (Fig. 1 and fig. S1). Each tilt series was acquired by rotating the sample around the x-axis with a tilt range from -62° to +62° in 4° increments. At each tilt angle, a focused x-ray beam was scanned over the sample with partial overlap between adjacent scan positions and a far-field diffraction pattern was recorded by a charge-coupled device camera at each scan position (see supplementary materials). The full data set consists of six tilt series with a total of 796,485 diffraction patterns.

The diffraction patterns were reconstructed using a regularized ptychographic iterative engine (*27*). After removing corrupt diffraction patterns and correcting the defocus at high tilt angles (figs. S2 and S3), a high-quality projection was obtained from the diffraction patterns at each tilt angle. Each pair of left- and right-circularly polarized projections was aligned and converted to the optical density for normalization. The sum of each pair of the oppositely-polarized projections produced three independent tilt series with different in-plane rotation angles. The scalar tomographic reconstruction was performed from the three tilt series of 96 projections by a real space iterative algorithm (see supplementary materials), which was able to optimize the reconstruction by iteratively refining the spatial and angular alignment of the projections. To validate the 3D scalar reconstruction and characterize the spatial resolution, we divided the 96 projections into two halves by choosing alternate projections. We conducted two independent reconstructions and calculated their Fourier shell correlation (fig. S4A), showing the consistency of the reconstructions with a resolution reaching the Nyquist limit. The 3D magnetization vector field was reconstructed from the difference of the left- and right-circularly polarized projections of the three tilt series by using least-squares optimization with gradient descent (see supplementary materials). Movie S1 shows the reconstructed 3D electron density and magnetization vector field of the ferromagnetic superlattice with a spatial resolution of 10 nm (fig. S4).

Next, we applied homotopy theory (*15*) to quantitatively analyze the experimentally determined 3D magnetization vector field and to identify non-collinear spin textures. In 3D magnetic systems, a topological point defect and the surrounding spin texture are referred to as a hedgehog. The topological charge, or hedgehog number within a bulk $\Omega$, follows the bulk-surface relationship (*26*),

$$Q = \int_\Omega \rho \, dxdydz = \int_{\partial\Omega} \boldsymbol{B}_e \cdot d\boldsymbol{S}, \qquad (1)$$

where $\rho = \frac{1}{8\pi}\epsilon^{ijk}\partial_i\boldsymbol{n}\cdot(\partial_j\boldsymbol{n}\times\partial_k\boldsymbol{n})$ is the hedgehog density, $\partial\Omega$ is the bounding surface, $\epsilon^{ijk}$ is the Levi-Civita symbol and $\boldsymbol{n}$ is the normalized magnetization vector field. $\boldsymbol{B}_e$ manifests as an emergent magnetic field acting on electrons and magnons moving through the magnetic texture, defined as $B_i = \frac{1}{8\pi}\epsilon^{ijk}\boldsymbol{n}\cdot(\partial_j\boldsymbol{n}\times\partial_k\boldsymbol{n})$. The right-hand side of Eq. (1) is commonly used to evaluate the skyrmions number in a 2D plane (*24, 25*), but can be generally applied to any 3D embedded surface (*28*). When the magnetization vector on the surface of a sphere enclosing a volume $\Omega$ covers the orientational parameter space exactly once, we have the hedgehog number $Q = \pm 1$, where +1 and -1 represent a hedgehog and an anti-hedgehog, respectively. According to Eq. (1), the hedgehog and the anti-hedgehog serve as a source and a sink of the emergent magnetic field $\boldsymbol{B}_e$. We also remark that Eq. (1) can be formulated on a lattice (*26*), which underpins our homotopic analysis on the experimental data.

To locate 3D topological defects from the magnetization vector field, we systematically searched for the local maxima of the hedgehog density ($\rho$). At each local maximum, we defined an enclosed surface with a radius of one voxel and calculated the hedgehog number using Eq. (1). Figure 2A and movie S2 show the 3D spatial distribution of 68 hedgehogs (red dots) and 70 anti-hedgehogs (blue dots) in the superlattice. We also correlated the location of these topological defects with the role of geometric confinement in the stabilization of hedgehogs (table S1). We found that hedgehogs and anti-hedgehogs are more abundant in octahedral than tetrahedral sites due to the larger volume of an octahedral site than of a tetrahedral site. Figure 2, B and D, shows a representative hedgehog and anti-hedgehog located in an octahedral and tetrahedral site, respectively. The 3D magnetization vector fields of the hedgehog and anti-hedgehog are shown in Fig. 2, C and E. The sign of the hedgehog number is not apparent from the 3D magnetization vector field, but can be unambiguously observed by computing the emergent magnetic field (fig. S5, A and B).

The localization of a total of 138 of hedgehogs and anti-hedgehogs allowed us to investigate their interactions in the ferromagnetic superlattice. According to the hedgehog confinement theory (*26*), the potential energy of a hedgehog and anti-hedgehog pair grows linearly with their separation when the exchange energy dominates, with all the emergent magnetic field lines emanating from a hedgehog ending at an anti-hedgehog. A non-negligible pair separation is an indication that competing interactions are driving the system into a frustrated state. Figure 3A shows a representative hedgehog and anti-hedgehog pair, where the emergent magnetic field lines were generated from the magnetization vector field. We observed that only part of the magnetic flux emanating from the hedgehog terminates at the anti-hedgehog, indicating that the emergent magnetic field lines are not completely confined. In comparison, the emergent magnetic field lines in a hedgehog/hedgehog and an anti-hedgehog/anti-hedgehog pair exhibit repulsive interactions (Fig. 3, B and C). Statistically, the distance of the hedgehogs/anti-hedgehog pairs was fit to be 18.3 ± 1.6 nm (Fig. 3D), while the hedgehog/hedgehog and anti-hedgehog/anti-hedgehog pairs were stabilized at comparatively larger distances of 36.1 ± 2.4 nm and 43.1 ± 2.0 nm (Fig. 3, E and F), respectively. The partial confinement of the emergent magnetic field lines (Fig. 3A) along with the observed hedgehog and anti-hedgehog pair distance suggests that additional factors competed with the exchange energy to stabilize the emergent magnetic monopoles in the superlattice, providing insight into understanding the superlattice as a magnetically frustrated system. Furthermore, the statistically significant difference in the nearest-neighbor distance between oppositely charged hedgehogs and similarly charged hedgehogs is consistent with theory (*26*), which further validates the quantitative nature of 3D soft x-ray vector ptychography.

According to the bulk-surface relationship of Eq. (1), the hedgehog number can be computed on an arbitrary surface. In the nickel superlattice, the silica nanospheres are magnetically inert voids and create 3D internal surfaces within the ferromagnetic network. To calculate the hedgehog number on these 3D surfaces, we performed a non-convex triangulation of the internal structure of the superlattice. The resulting facets were grouped into individual void surfaces using a community-clustering technique in network analysis (*29*). Due to the finite thickness of the sample, the majority of the magnetic voids are not fully closed, meaning that they cannot have integer hedgehog numbers. In accordance with the literature (*30*), we defined any void surface with $Q \geq 0.9$ as a virtual hedgehog and $Q \leq -0.9$ as a virtual anti-hedgehog. Figure 2A and movie S2 show the distribution of 8 virtual hedgehogs (red) and 11 virtual anti-hedgehogs (blue) in the ferromagnetic superlattice. A representative virtual hedgehog and anti-hedgehog with $Q = 1.01$ and -1 are shown in Fig. 4, A and B, respectively. The 3D magnetization vector field of the virtual hedgehog and anti-hedgehog were mapped onto a 2D plane to produce two stereographic projections (Fig. 4, C and D). For the virtual hedgehog, most spins point down in the center and point up at the boundary, while for the virtual anti-hedgehog, most spins point up in the center and point down at the boundary. The emergent magnetic field was also calculated for the 3D virtual topological defects (fig. S5, C and D). These virtual topological defects indicate 3D spin textures consistent with emergent magnetic monopoles residing at the geometric centers of the magnetic voids, which is a clear manifestation of the bulk-surface correspondence.

The existence of topological defects in a system only comprised of nickel is surprising, as it does not possess strong anisotropy or Dzyaloshinskii-Moriya interaction as seen in other materials that usually support topological defects, e.g., non-centrosymmetric lattices and magnetic / heavy-metal multilayers (*22–25, 31*). However, surface curvature can stabilize magnetic solitons through an effective Dzyaloshinskii-Moriya interaction (*14, 32*). The nickel superlattice is a 3D generalization of a Kagome lattice (*33*), where the magnetic material was infiltrated into a close-packed non-magnetic lattice, resulting in magnetic frustration. Our experimental results provide direct evidence for the existence of isolated magnetic hedgehogs and virtual hedgehogs in the ferromagnetic superlattice, showing that the competing energies in the frustrated superlattice can unlock the richness and functionality of nickel that was previously unseen. Our observations also give insight into the confinement-deconfinement transition of hedgehogs (*26*). Thus, this work could open the door to use magnetically frustrated superlattices as a platform to study hedgehog interactions, dynamics, and confinement-deconfinement transition as well as exploit them for spin transport applications. Furthermore, the spatial resolution of this experimental method can be readily extended to the nanometer scale by increasing the incident coherent flux or the data acquisition time. We expect that nanoscale soft x-ray vector ptychography can be applied to quantitatively image 3D topological spin textures in a wide range of magnetic materials as well as the 3D vector fields in other disciplines.

**ACKNOWLEDGMENTS**: We thank Rafal Dunin-Borkowski and Jong E. Han for stimulating discussions. **Funding**: This work was primarily supported by STROBE: a National Science Foundation Science and Technology Center under award DMR1548924. J.M. and A. R.

acknowledge partial support by the US Department of Energy, Office of Science, Basic Energy Sciences, Division of Materials Sciences and Engineering under award number DE-SC0010378 for the development of vector ptychography. M.M. and H.K. acknowledge partial support by the US Department of Energy, Office of Science, Basic Energy Sciences X-Ray Scattering Program Award DE-SC0002002 and DARPA TEE Award No. D18AC00017 for the data acquisition and analysis. Y.T. and J.Z. were supported by the U.S. Department of Energy, Office of Basic Energy Sciences under Grant No. DE-SC0012190. **Author contributions**: J.M. directed the project and M.M.M. suggested the sample; A.J.G, J.V.B, P.M., T.E.M., C.-T.L. and S.Y. synthesized and fabricated the sample; A.R., C.-T.L., Y.H.L., E.E.C.S., S.R., X.L., C.S.B., R.M.K., Y.S.Y., D.A.S, H.C.K., M.M.M. and J.M. planed and/or performed the experiments; M.P., A.R., S.J.O. and J.M. developed the scalar and vector tomography algorithms; A.R. and J.M. reconstructed the 3D magnetization vector field; A.R., E.I. J.Z. and J.M. analyzed the data with input from M.M.M., Y.T., C.-T.L. and V.H.C.; A.R., J.M., E.I. and J.Z. wrote the manuscript with input from M.M.M., Y.T., C.-T.L., S.Y., E.E.C.S. and T.E.M. The x-ray ptychography experiments were performed at COSMIC used resources of the Advanced Light Source, which is a DOE Office of Science User Facility under contract no. DE-AC02- 05CH11231. **Competing interests**: The authors declare no competing financial interests. **Data and materials availability**: All data are available in the main text or the supplementary materials.

**FIGURES**

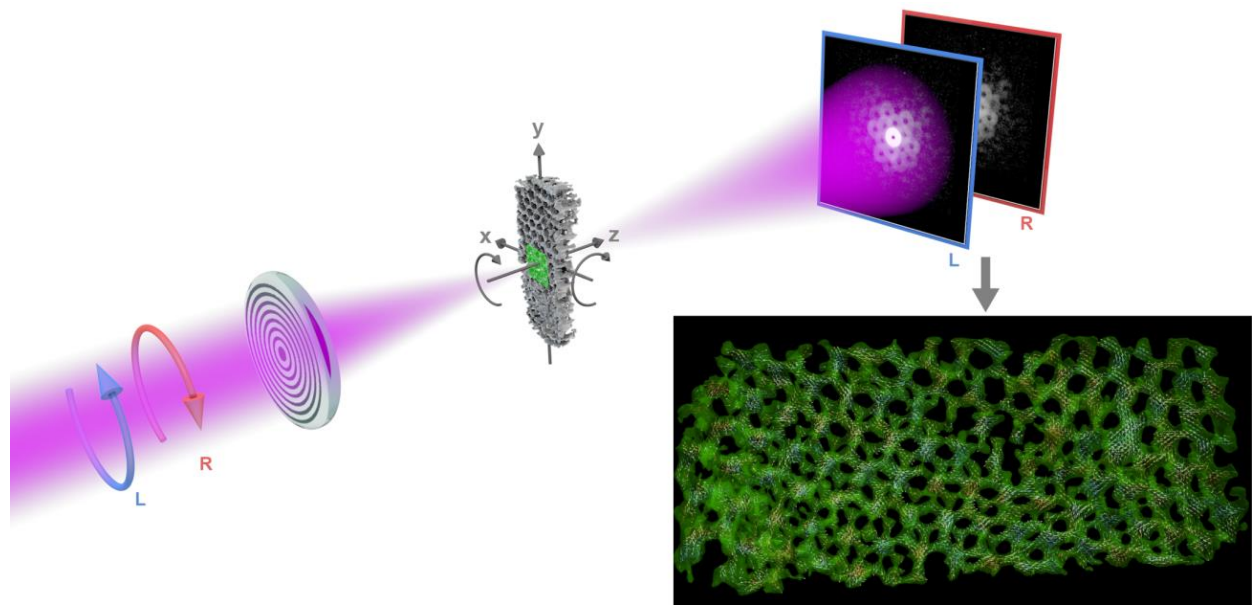

**Fig. 1. Experimental schematic of 3D soft x-ray vector ptychography.** Left and right circularly polarized x-rays (pink) were focused onto a ferromagnetic superlattice sample (center) with green circles indicating the partially overlapped scan positions. The sample was tilted around the x- and z-axis and diffraction patterns were collected by a charge-coupled device camera. The lower right structure shows the 3D electron density (green) and magnetization vector field (arrows) of the superlattice reconstructed from the diffraction patterns (see movie S1 for more detail).

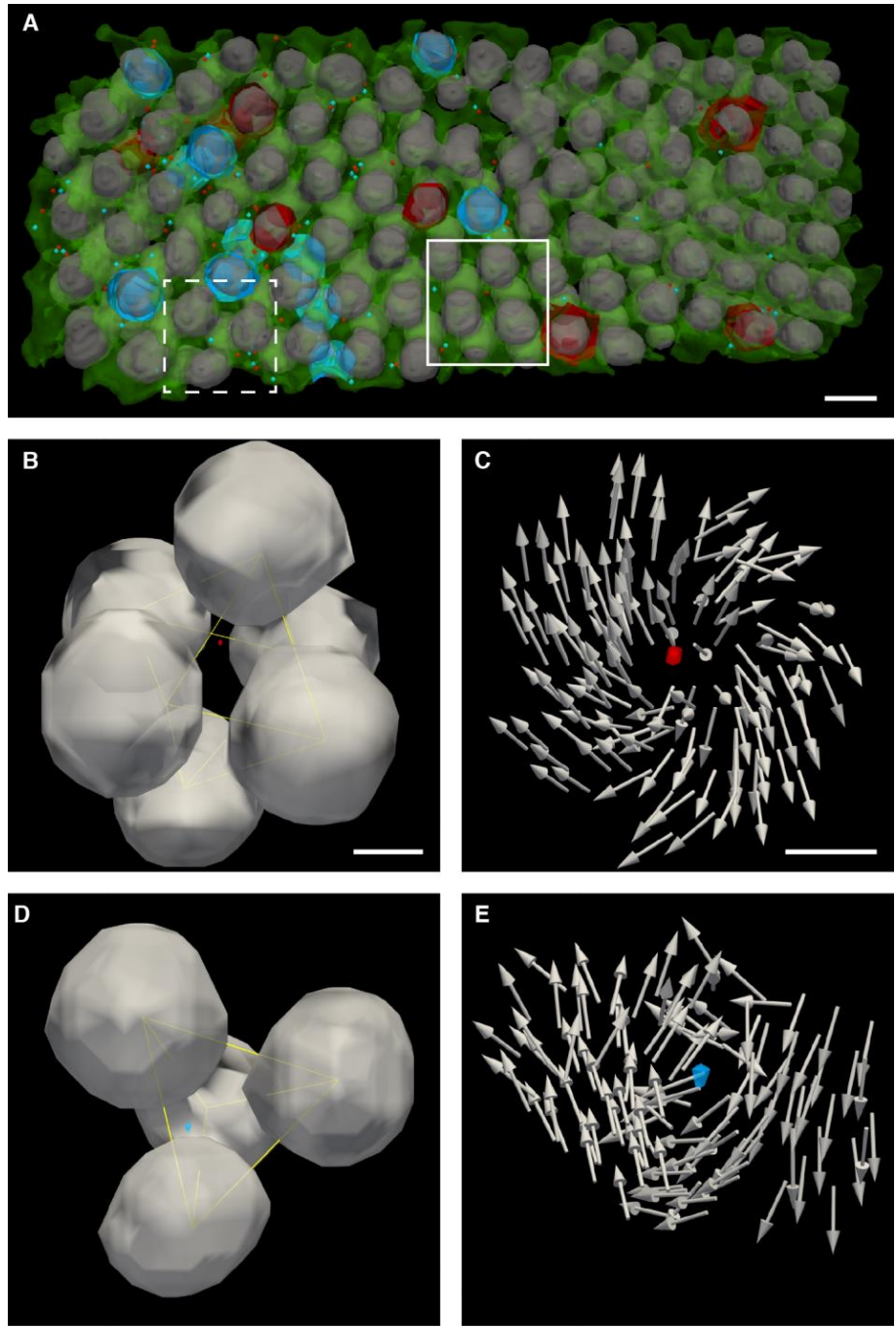

**Fig. 2. Quantitative 3D characterization of hedgehogs and anti-hedgehogs in the ferromagnetic superlattice.** (**A**) 3D spatial distribution of 68 hedgehogs (red dots) and 70 anti-hedgehogs (blue dots) in the superlattice, where the surfaces of the magnetic voids in red and blue represent virtual hedgehogs and virtual anti-hedgehogs, respectively. The silica nanospheres are rendered as gray iso-surfaces The solid and dashed squares mark the region of interest shown in

(B) and (D), respectively. (**B** and **C**) The location and 3D spin texture of a hedgehog within a tetrahedral site of the face-centered cubic superlattice. (**D** and **E**) The location and 3D spin texture of an anti-hedgehog within an octahedral site. Scale bars in (A-C) are 60, 25 and 10 nm, respectively.

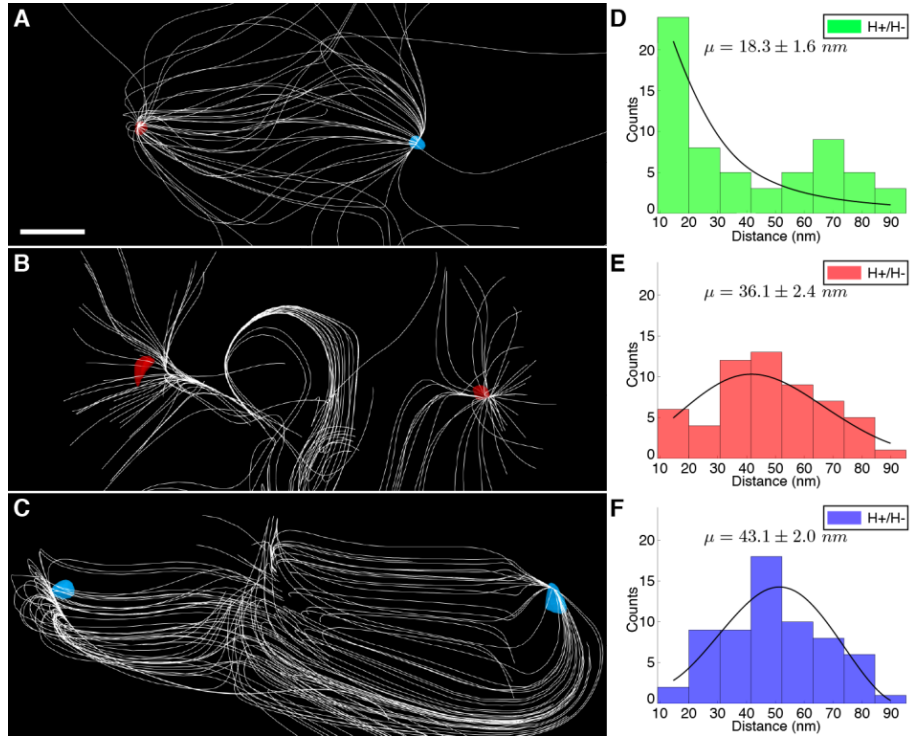

**Fig. 3. Quantification of the interactions of the emergent magnetic monopoles in the ferromagnetic superlattice.** Representative hedgehog/anti-hedgehog (**A**), hedgehog/hedgehog (**B**) and anti-hedgehog/anti-hedgehog (**C**) pair and their emergent magnetic field lines. As the emergent magnetic field lines are continuous, we obtained them from interpolated 3D magnetization vector field. Note that the interpolation is for display purposes. All the quantitative analysis using Eq. (1) in this work was performed from non-interpolated experimental data. Histograms of the nearest-neighbor distance for the hedgehog/anti-hedgehog (**D**), hedgehog/hedgehog (**E**) and anti-hedgehog/anti-hedgehog (**F**) pair. The three histograms were fit to a generalized extreme value distribution, producing three curves in (D-F), where $\mu$ represents

the center of each fit and the standard error was determined from the fit's confidence interval. Scale bar, 5 nm.

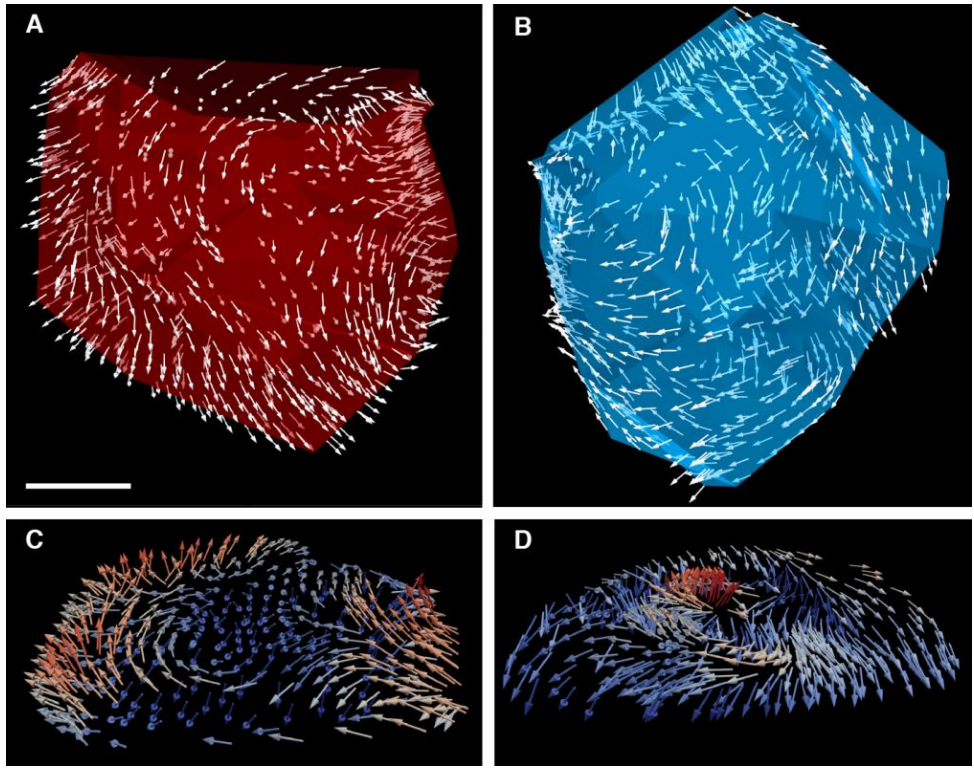

**Fig. 4. Representative virtual hedgehog and anti-hedgehog in the ferromagnetic superlattice.** **(A)** and **(B)** A virtual hedgehog and anti-hedgehog with $Q = 1.01$ and -1, respectively, where the arrows indicate the 3D magnetization vector field. **(C)** and **(D)** Stereographic projections of the virtual hedgehog and anti-hedgehog shown in (A) and (B), respectively, where the colors of the arrows represents the z-component of the spin with red corresponding to pointing up (+z) and blue pointing down (-z). Scale bar, 15 nm.

# Supplementary materials

## Materials and Methods

### Sample synthesis and preparation

The 3D ferromagnetic superlattice was synthesized by infiltrating interconnected voids of a silica nanoparticle template using confined chemical fluid deposition. Monodisperse silica nanoparticles of ~69 nm diameter (standard deviation < 5%) were synthesized using a liquid-phase method (*34*). The evaporation-assisted vertical deposition technique was used to assemble these particles onto silicon substrate (*35*). Briefly, 3 cm x 1 cm silicon wafers were placed at a ~30° angle in open plastic vials containing 10x dilute solution of the as-synthesized particles. The vials were left undisturbed for two weeks in an oven maintained at 40° C at 80% relative humidity. The resulting films that were used as the template for nickel infiltration contained silica particles arranged in a mixture of close-packed structures and had thicknesses ranging from 240 nm – 850 nm depending on the vertical position of the silicon substrate (*36*).

   The infiltration of nickel within the template voids was performed using confined chemical fluid deposition (*20*). In short, the template was spatially confined using a 250 µm thick U-shaped titanium spacer and placed within a custom-built reactor made of parts from High Pressure Equipment Company, McMaster, and Swagelok. Bis(cyclopentadienyl) nickel (II) was loaded into the reactor in a Vacuum Atmospheres argon glovebox. The reactor was pressurized with Praxair 4.0 Industrial Grade carbon dioxide using a custom-made manual pump and heated to 70°C for 8 hours at a pressure of around 13.8 MPa to dissolve the precursor powder into the supercritical carbon dioxide. A separate gas reservoir was loaded with Praxair 5.0 ultra-high purity hydrogen using a Newport Scientific Two Stage 207 MPa Diaphragm Pump and was connected to the reactor. The hydrogen was added to the reactor to a final reactor pressure of 42.7 MPa and the deposition proceeded at 100°C for 10 hours. The interstitial voids between the nanospheres of the template were then infiltrated with nickel, forming a superlattice. An overfilled nickel film over the superlattice and template resulting from the deposition process was milled using a Leica EM TIC 3x Argon ion beam milling system at 3 degrees and 3kV.

   To prepare the sample with the correct geometry for our 3D vector ptychography experiments, we lifted out a portion of the sample from the bulk superlattice on a silicon substrate and thinned the sample using a focused ion beam (FIB, FEI Nova 600 NanoLab DualBeam), which was equipped with a field emission scanning electron microscope and a scanning gallium ion beam. The FIB prepared sample was mounted on a 3-mm TEM grid (Omniprobe, 3 posts copper lift-out grid), where the central post was also trimmed by FIB milling to increase the tilt range. The sample mounted on the TEM grid was examined by the scanning electron microscope and an optical microscope, and then manually glued on a 3-mm copper ring using a silver paste (fig. S1). The sample fabricated by this process consists of 1-2 layers of nanospheres and can be manually rotated in-plane for the 3D vector ptychography experiment.

### The 3D soft x-ray vector ptychography experiment

The experiment was conducted at the COSMIC beam line at the Advanced Light Source, Lawrence Berkeley National Lab (*37*). Figure 1 in the main text shows the experimental schematic of 3D soft x-ray vector ptychography. An elliptical polarization undulator allowed us to generate circularly polarized x-rays of left- and right-helicity and achieve differential contrast enhancement of the magnetic signal. The incident photon energy was tuned to 856 eV, slightly above the nickel L$_3$ edge, to maximize the absorption contrast. The polarized beam was focused onto the sample by a Fresnel zone plate with an outer width of 45 nm. A total of six tilt series were acquired from the sample with left- and right-circularly polarized x-rays at three in-plane rotation angles (0°, 120° and 240°). At each in-plane rotation angle, the sample was tilted from -62° to +62° in 4° increments. At each tilt angle, the focused beam was raster-scanned across the sample in 40 nm steps. Diffraction patterns were collected using both left and right circularly polarized x-rays. A far-field CCD camera was used to record the diffraction patterns at each scan position. Initial reconstructions were performed on-site in real time using a GPU-based ptychography reconstruction algorithm (*38*).

**Data processing and ptychographic reconstructions**

A very small number of corrupt diffraction patterns, most commonly caused by detector readout malfunction or unstable beam flux, resulted in a global degradation of the reconstructed object wave-front through the coupling of the probe and object. We used the following procedure to automatically detect and remove the corrupt diffraction patterns to achieve the high-quality reconstructions. The high-angle diffraction intensity at each scan position was integrated to produce a low-resolution map at every ptychography scan. Local maxima in the magnitude of the gradient of this map were used to identify and remove bad frames as a pre-reconstruction data processing step (fig. S2). In addition, a phase unwrapping constraint was enforced on the probe as part of each macro-iteration of the reconstruction to deal with defocus at high tilt angles (fig. S3). The final ptychographic reconstructions were performed using the regularized ptychographic iterative engine (*27*) with a modification to address defocus at high tilt. From the reconstructed complex-valued exit wave, the absorption component was used as the magnetic contrast (*39*) and the two oppositely-polarized projections at each tilt angle were aligned using a feature-based image registration package in MATLAB. The projections were converted to optical density (*40*) to normalize any small temporal and polarization-based fluctuations of the beam intensity. In each projection, background subtraction was performed by numerically evaluating Laplace's equation, using the region exterior to the sample as the boundary condition (*41*). We found that this method outperforms simple constant background subtraction, as it can accommodate local variations.

**The scalar tomography reconstruction**

The relationship between charge and magnetic scattering (*42*),

$$f = f^c \pm if^m \hat{z} \cdot \boldsymbol{m} ,  \qquad (2)$$

was used to generate a set of scalar and vector projections corresponding to the charge and magnetic scattering, where $f^c$ and $f^m$ are the charge and magnetic scattering factor, respectively, $\hat{z}$ is the x-ray propagation direction, and $\boldsymbol{m}$ is the magnetization vector. The sum of each pair of the oppositely-polarized projections produced three independent tilt series with different in-plane rotation angles. The scalar projections of each tilt series were first roughly aligned with cross-correlation, then more accurately aligned using the center-of-mass and common line method (*43*). The aligned tilt series was reconstructed by a real space iterative reconstruction (RESIRE) algorithm (*44*), which was able to iteratively perform angular and spatial refinement to adjust any remaining small alignment errors. From the three independent reconstructions, transformation matrices were computed to align the three tilt series of 96 projections to a global coordinate system. The three aligned tilt series were collectively reconstructed by RESIRE using the same angular and spatial refinement procedure, which produced the final scalar tomography reconstruction. The transformation matrices obtained from the scalar tomography were used for the vector tomography reconstruction.

**The vector tomography reconstruction**

The 3D vector magnetization field was reconstructed by taking the difference of the left- and right-circularly polarized projections of the six experimental tilt series, producing three independent tilt series with the magnetic contrast. The vector tomography algorithm is modeled as a least squares optimization problem and is solved directly by gradient descent. The least squares problem is given as,

$$min_{O_1,O_2,O_3} f(O_1,O_2,O_3) = \sum_{i=1}^{N} || \alpha_i \Pi_i O_1 + \beta_i \Pi_i O_2 + \gamma_i \Pi_i O_3 - b_i ||^2$$

$$= \sum_{i=1}^{N} || \Pi_i(\alpha_i O_1 + \beta_i O_2 + \gamma_i O_3) - b_i ||^2 \quad , \quad (3)$$

where $O_1$, $O_2$, $O_3$ are the three components of the vector field to be reconstructed, $N$ is the number of the projections of the three tilt series, $\Pi_i$ is the projection operator with respect to the Euler angle set $\{\phi_i, \theta_i, \psi_i\}$, and $b_i$ is the experimentally measured projection. $\{\alpha_i, \beta_i, \gamma_i\}$ are the coefficient set with respect to the projection operator and are related to the corresponding Euler angle set by,

$$\alpha_i = \sin \theta_i \cos \phi_i, \quad \beta_i = \sin \theta_i \sin \phi_i, \quad \alpha_i = \cos \theta_i \quad . \quad (4)$$

The least square problem is solved via gradient descent and the gradients are computed by,

$$\nabla_{O_1} f(O_1,O_2,O_3) = \sum_{i=1}^{N} \alpha_i \Pi_i^T \Pi_i(\alpha_i O_1 + \beta_i O_2 + \gamma_i O_3)$$

$$\nabla_{O_2} f(O_1,O_2,O_3) = \sum_{i=1}^{N} \beta_i \Pi_i^T \Pi_i(\alpha_i O_1 + \beta_i O_2 + \gamma_i O_3) \quad . \quad (5)$$

$$\nabla_{O_3} f(O_1,O_2,O_3) = \sum_{i=1}^{N} \gamma_i \Pi_i^T \Pi_i(\alpha_i O_1 + \beta_i O_2 + \gamma_i O_3)$$

The $(j+1)^{th}$ iteration of the algorithm is updated as,

$$O_1^{j+1} = O_1^j - t \nabla_{O_1} f(O_1,O_2,O_3) = O_1^j - t \sum_{i=1}^{N} \alpha_i \Pi_i^T \Pi_i(\alpha_i O_1^j + \beta_i O_2^j + \gamma_i O_3^j)$$

$$O_2^{j+1} = O_2^j - t \nabla_{O_2} f(O_1,O_2,O_3) = O_2^j - t \sum_{i=1}^{N} \beta_i \Pi_i^T \Pi_i(\alpha_i O_1^j + \beta_i O_2^j + \gamma_i O_3^j) \quad , \quad (6)$$

$$O_3^{j+1} = O_3^j - t\,\nabla_{O_3} f(O_1, O_2, O_3) = O_3^j - t \sum_{i=1}^{N} \gamma_i \Pi_i^T \Pi_i (\alpha_i O_1^j + \beta_i O_2^j + \gamma_i O_3^j)$$

where $t$ is the step size. For a given tilt angle set $\{\phi_i, \theta_i, \psi_i\}$, the forward projection of a 3D object is computed using the Fourier slice theorem, while the back projection is implemented by linear interpolation.

To validate the vector tomography algorithm, we simulated 3D topological spin textures and calculated their diffraction patterns based on the experiential parameters. After adding noise to the diffraction patterns, we performed ptychographic reconstructions to generate projections. Using the vector tomography algorithm, we were able to reconstruct the 3D topological spin textures from the projections. After validating the vector tomography algorithm using simulated data, we applied it to reconstruct the 3D vector magnetization field of the ferromagnetic superlattice from the experimentally measured tilt series.

**Quantification of the spatial resolution**

We quantified the spatial resolution using three independent methods. First, in the scalar tomography reconstruction, we divided the 96 projections of three tilt series into two halves by choosing alternate projections. We conducted two independent reconstructions and calculated their Fourier shell correlation (FSC) (fig. S4A). According to the cutoff of FSC = 0.143, a criterion commonly used in cryo-electron microscopy (*45*), we concluded that a spatial resolution of 10 nm was achieved in the scalar tomography reconstruction. Second, by taking the difference of the experimental diffraction patterns measured with left- and right-circularly polarized soft x-rays, we calculated the power spectral density of the superlattice's magnetic contrast (fig. S4B), showing that the magnetic signal extends to a spatial resolution of 10 nm. Third, we calculated the hedgehog number of a reconstructed hedgehog/anti-hedgehog pair with a separation distance of 10 nm (fig. S4, C-E). The hedgehog number of the pair is $Q = 0$, while the hedgehog number of the individual anti-hedgehog and hedgehog is $Q = -1$ and $+1$, respectively. This analysis shows that a spatial resolution of 10 nm was achieved in the vector tomography reconstruction.

**Supplementary Figures**

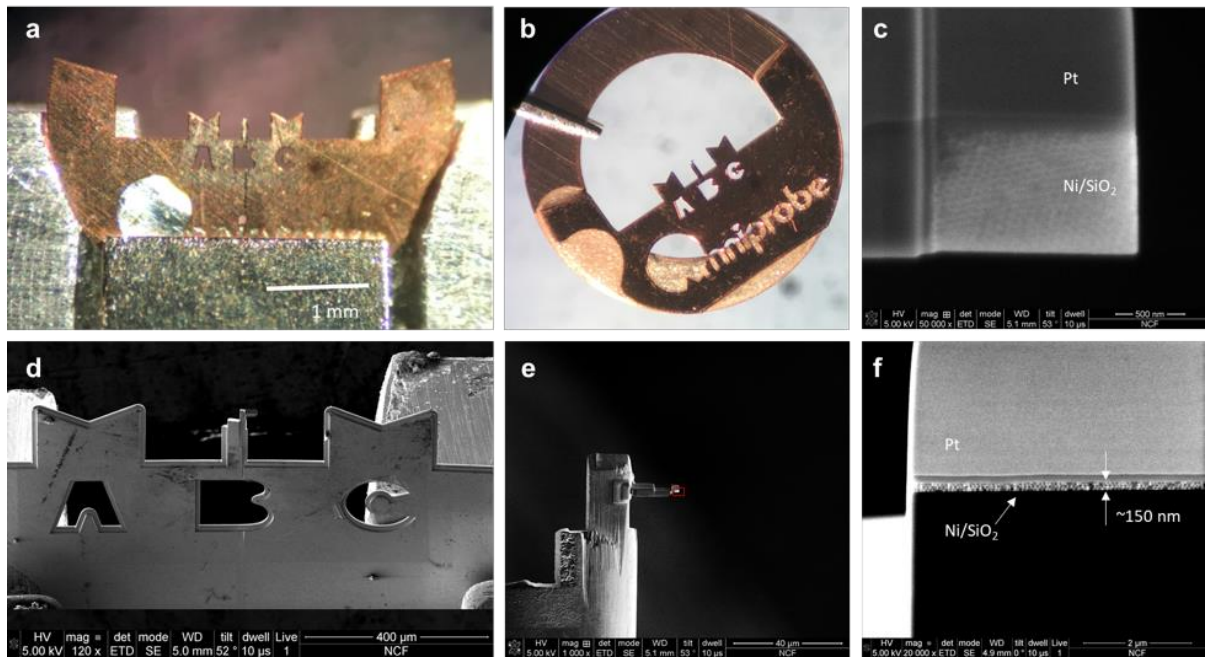

**Fig. S1. Sample preparation.** (**A** and **B**) Optical microscopy images of the superlattice sample, prepared by focused ion beam milling. The sample was mounted on a 3-mm transmission electron microscopy grid and glued on a copper ring (B). (**C-F**) Scanning electron microscopy images of the sample and its mounting geometry. The mounting geometry is important for the 3D soft x-ray vector ptychography experiment with three in-place rotation angles. The sample is 150 nm thick as shown in a side view image (F).

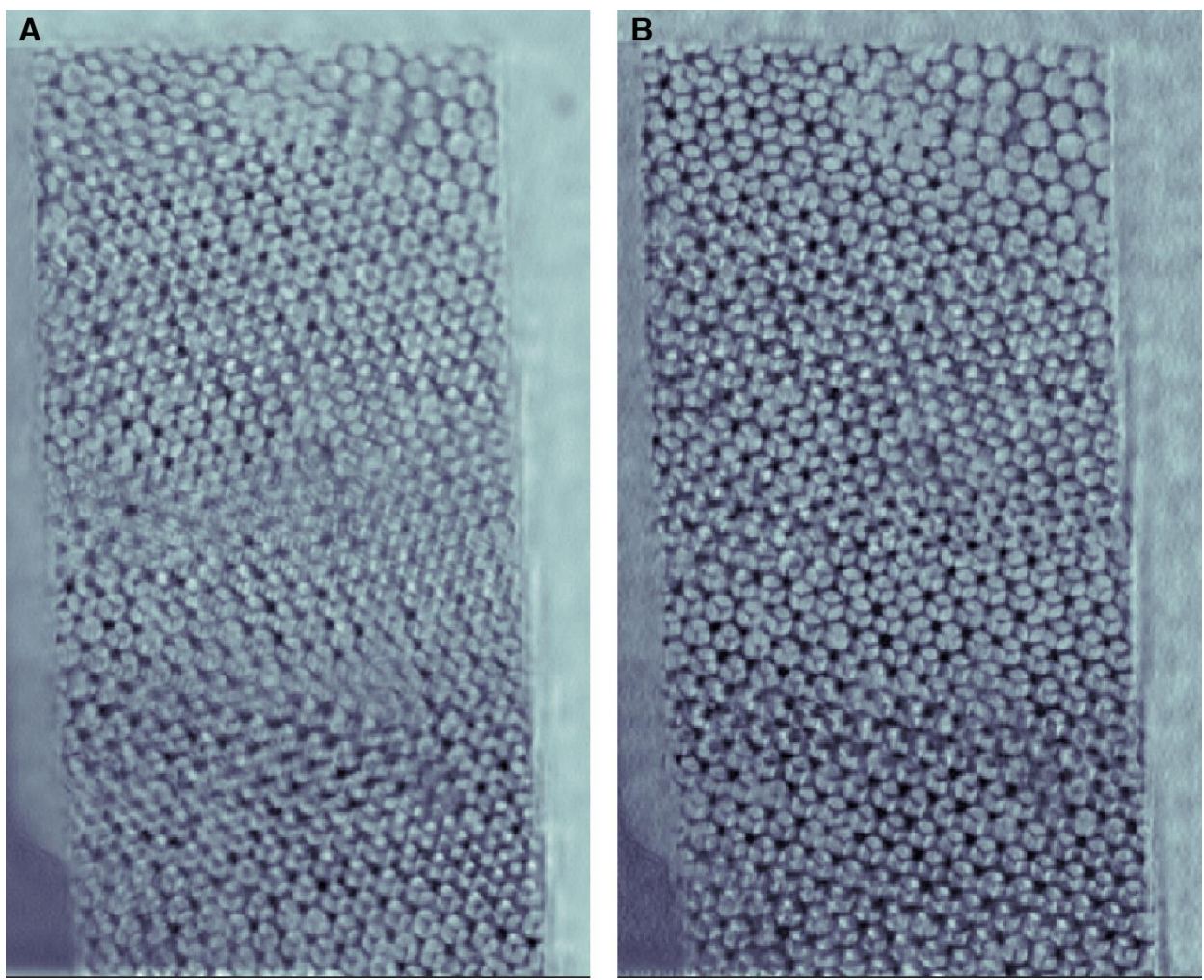

**Fig. S2. Ptychography reconstructions before and after the removal of a very small number of corrupt diffraction patterns**. (**A**) A ptychography reconstruction of a projection with a very small number of corrupt diffraction patterns, which were caused by detector readout malfunction or unstable x-ray flux, and produced reconstruction artifacts. (**B**) The same reconstructed projection after the removal of the corrupt diffraction patterns.

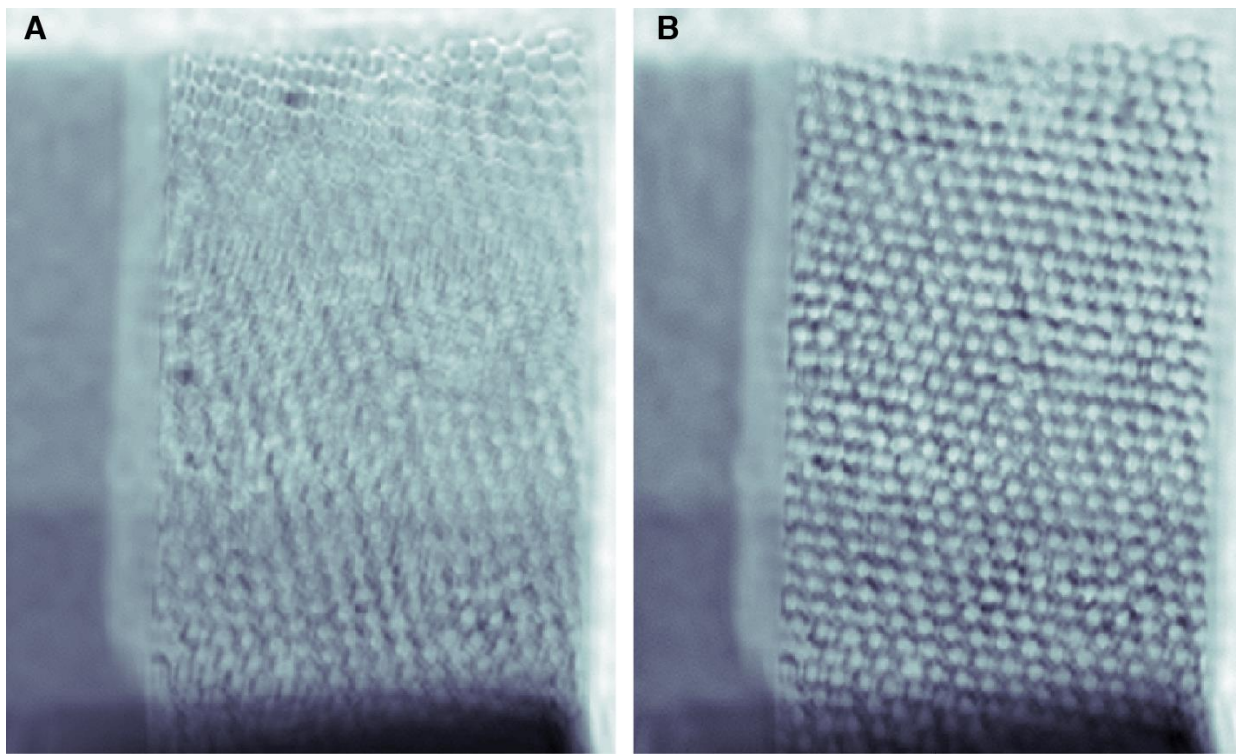

**Fig. S3. Ptychography reconstructions before and after defocus correction. (A)** The reconstruction of a high-tilt projection, in which the artefacts were induced by the varied defocus values at different sample regions. **(B)** The same reconstructed projection after defocus correction.

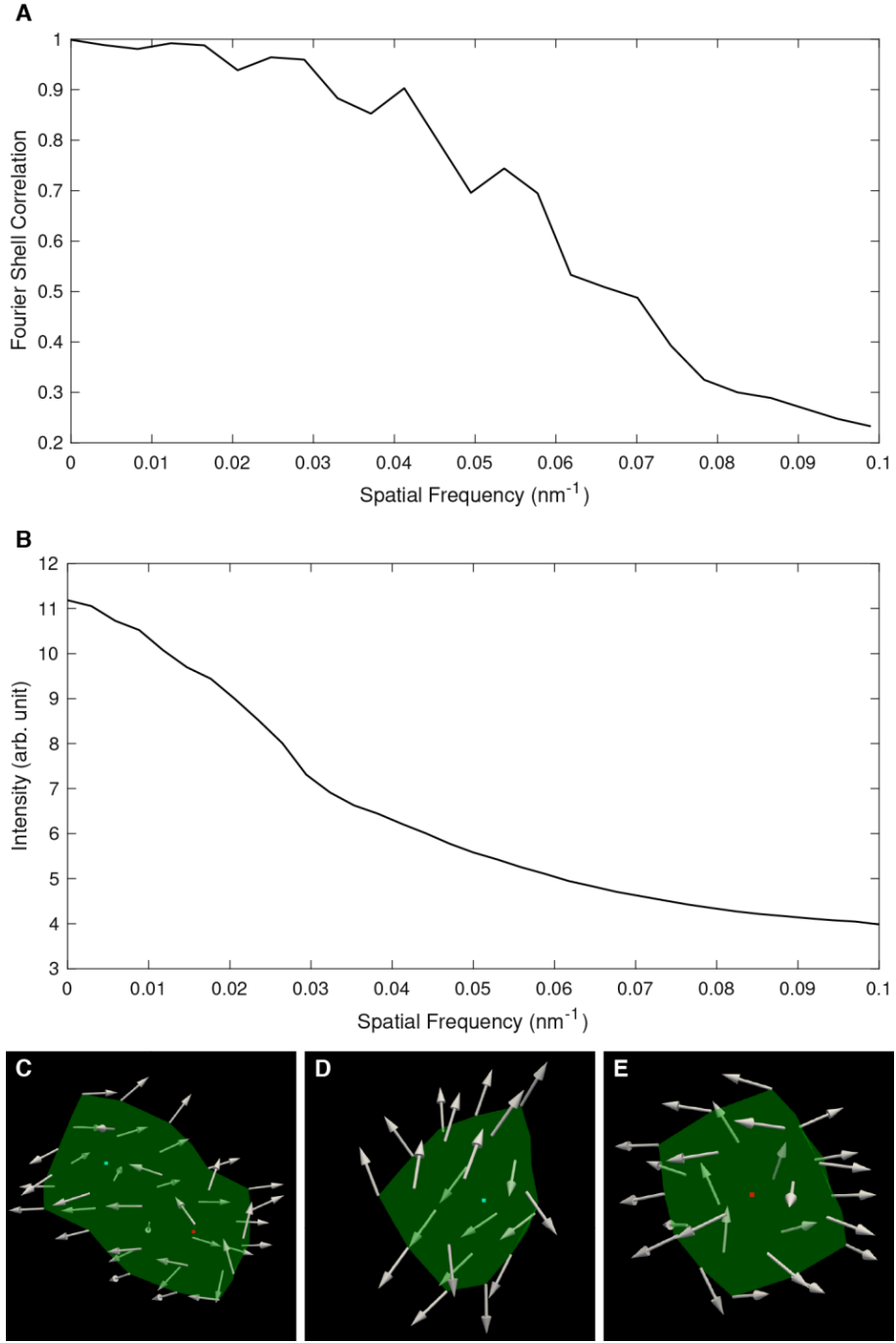

**Fig. S4. Quantification of the spatial resolution.** (**A**) The Fourier shell correlation (FSC) of two independent scalar tomography reconstructions. According to a criterion of FSC = 0.143 (*45*), a spatial resolution of 10 nm was achieved, which corresponds to a spatial frequency of 0.1 nm$^{-1}$. (**B**) Power spectral density of the magnetic contrast from all the diffraction patterns, showing that the signal extends to a spatial resolution of 10 nm. A reconstructed hedgehog/anti-hedgehog pair with a separation distance of 10 nm. The hedgehog number was evaluated for the pair with $Q = 0$ (**C**), for the anti-hedgehog with $Q = -1$ (**D**) and for the hedgehog with $Q = +1$ (**E**).

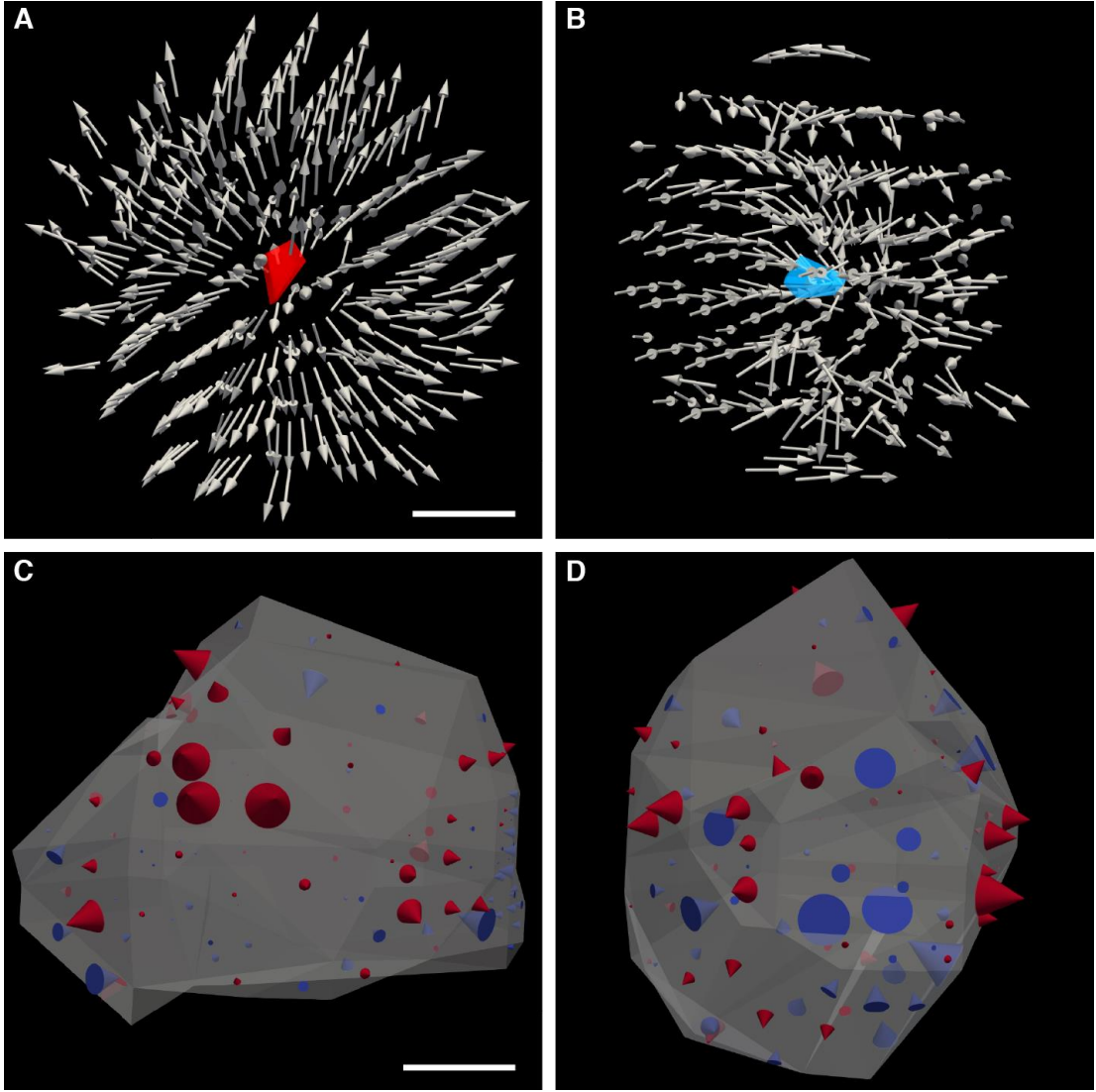

**Fig. S5. The emergent magnetic field of hedgehogs and virtual hedgehogs. (A)** and **(B)** The emergent magnetic field of the hedgehog and anti-hedgehog shown in Fig. 2, C and E, respectively, where interpolation was used to display the $\mathbf{B_e}$ field. The vector plots indicate that the hedgehog and the anti-hedgehog form a source and a sink of the emergent magnetic field. **(C)** and **(D)** The emergent magnetic field of a virtual hedgehog and virtual anti-hedgehog, respectively. The red and blue cones represent outflow and inflow of the emergent magnetic field, respectively, while the cone size indicates the total emergent flux through the facet. Note that while there is both outflow and inflow of the emergent magnetic field in each case, the net flow corresponds to a source and sink, respectively. The scale bars in (A) and (C) are 5 nm and 15 nm, respectively.

| Location | Number of hedgehogs | Number of anti-hedgehogs | Total number |
|---|---|---|---|
| Octahedral sites | 44 | 46 | 90 |
| Tetrahedral sites & the space between the two sites | 24 | 24 | 48 |

**Table S1**: **Locations of hedgehogs and anti-hedgehogs in the ferromagnetic superlattice.**

**Movie S1**. 3D scalar (green) and vector (arrow) reconstructions of the ferromagnetic superlattice. The global view of the 3D magnetization vector field zooms in to show a pair of hedgehog/anti-hedgehog, hedgehog/hedgehog and anti-hedgehog/anti-hedgehog, corresponding to Figs. 3, A-C, respectively. In each zoomed-in field of view, the global field fades away and the local magnetization vector field around each hedgehog or anti-hedgehog is given by gray arrows. The field lines follow the emergent $\mathbf{B_e}$ field.

**Movie S2**. 3D scalar reconstruction of the ferromagnetic superlattice (green), where the locations of hedgehogs and anti-hedgehogs are marked with red and blue dots, respectively. Virtual hedgehogs and anti- hedgehogs with $Q \geq 0.9$ and $Q \leq -0.9$ are labelled with red and blue triangulated surfaces, respectively. The silica nanospheres are rendered as gray iso-surfaces.